\documentclass[12pt]{article}
\usepackage{epsfig}

\def\Title#1{\begin{center} {\Large {\bf #1} } \end{center}}

\begin{document}

\Title{Quark Stars and Color Superconductivity:\\ A GRB connection ?}

\bigskip\bigskip

%+\addtocontents{toc}{{\it D. Reggiano}}
%+\label{ReggianoStart}

\begin{raggedright}

{\it Rachid Ouyed\index{Ouyed, R.}\\
Nordic Institute for Theoretical Physics (NORDITA)\\
Blegdamsvej 17\\
DK-2100 Copenhagen \O, Denmark}
\bigskip\bigskip
\end{raggedright}

\abstract

At this conference many interesting talks
were presented on the plausible existence of Quark
Stars. Other talks dealt with  the
exotic new phases  of quark matter 
at very high density. Here, I show how combining these
two elements might
offer a new way of tackling the Gamma Ray Burst 
puzzle. 
\vskip0.5cm

\section{Introduction}

It is widely accepted that
the most conventional interpretation of the observed Gamma-ray bursts (GRBs)
result from the conversion of the kinetic energy
of ultra-relativistic particles to radiation in an
optically thin region \cite{kouveliotou95,
kulkarni99,piran99a,piran99b}. The particles being
accelerated by a fireball mechanism (or explosion of
radiation) taking place
near an unknown central engine \cite{goodman86,shemi90,paczynski90}. 
The first challenge is to conceive of circumstances that would
create a sufficiently energetic fireball.
In the model presented in this talk, the approach  
is to make use of intrinsic
properties of quark stars (where
exotic phases of quark matter come into play) to account for the
fireball.  

Quark matter at very high density  is
expected to behave as a color superconductor (see
Figure 1). 
A novel feature of such a phase (in the
2-flavor case; hereafter 2SC) is the generation of  glueball like particles
(hadrons made of gluons) which as demonstrated
in \cite{ouyed01a} immediately decay into photons. 
If  2SC  sets in at the surface
of a quark star the glueball decay
becomes a natural mechanism for a fireball
generation; a mechanism which is
fundamentally different from models
where the fireball is generated via
a collapse 
\cite{blandford77,ruffert99,janka99}
or conversion (of neutron star
to quark star \cite{olinto87,cheng96,bombaci00}) processes.
I will then show how and why quark stars 
might constitute new 
candidates for GRB inner engines 
\cite{ouyed01b}.

\section{Quark stars and the 2SC phase}

\begin{figure}[t]
\begin{center}
\epsfig{file=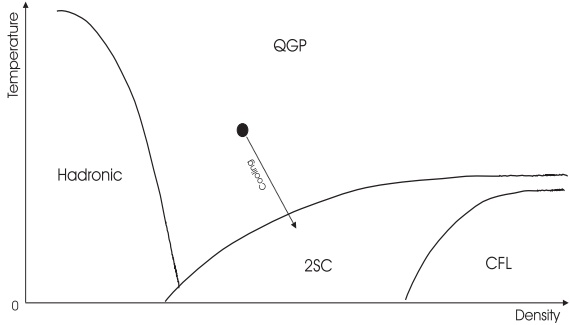,width=3.5in,height=3.5in,angle=-90}
\vskip -1.0in
\caption{A schematic representation of
a possible QCD phase diagram (see Alford in this volume).
The arrow depicts a plausible cooling path of
a quark star leading to the onset of color superconductivity
in its surface.}
\label{figure1}
\end{center}
\end{figure}

We first assume that quark stars exists in nature (further discussed in \S 7.2; see
also Heiselberg and Bombaci in this volume)
and constitutes the first major assumption in our
model.

\subsection{Hot Quark stars}

We are concerned with quark stars
born with temperatures above $T_{c}$ 
(the critical temperature
above which thermal fluctuations will
wash out the superconductive state).
We shall refer to these stars as ``hot'' quark stars
(HQSs) in order to avoid any confusion with strange stars
which are conjectured to exist even at zero pressure if strange matter
is the absolute ground state of strong interacting matter rather than
iron \cite{bodmer71,witten84,haensel86,alcock86,dey98}.

Among the features of HQSs relevant to our
model \cite{alcock86,farhi84,glendenning92,glendenning97}:

i) The ``surface" of a HQS is very
different from the surface of a neutron star, or any other type of stars.
Because it is bound by the strong force, the density at the
surface changes abruptly from zero to $\rho_{HQS}$. The abrupt change 
(the thickness of the quark surface) occurs within
 about 1 fm, which is a typical strong
interaction length scale.

ii) The electrons being
bound to the quark matter by the electro-magnetic
 interaction and not by the strong
force, are able to move freely across the quark surface 
extending up to $\sim
10^3$ fm above the surface of the star.
Associated with this electron layer is a
strong electric field ($5\times 10^{17}$ V/cm)- higher than the critical
value ($1.3\times 10^{16}$ V/cm) to make the vacuum region unstable
 to spontaneously create  $(e^{+},e^{-})$ pairs.

iii) The presence of normal matter (a crust made of ions)
at the surface of the quark star is subject to the 
enormous electric dipole. The strong positive Coulomb
 barrier prevents atomic
nuclei bound in the nuclear crust from coming into
direct contact with the quark
core. The crust \cite{baym71} is suspended above the
vacuum region. 

iv) One can show that the maximum
mass of the crust cannot exceed $M_{crust}\simeq 5\times 10^{-5}M_{\odot}$
set by the requirement that if the density in the inner
crust is above the neutron drip
density ($\rho_{drip}\simeq 4.3\times 10^{11}$ g/cc),
free neutrons will gravitate to the surface of
the HQS and be converted to quark matter.
This is due to the fact
that neutrons can easily penetrate the
Coulomb barrier and are
readily absorbed. 

\subsection{2SC and Light GlueBalls (LGBs)}

The 2SC phase 
is characterized by five out of the
eight gluons acquiring mass. 
The 3 massless gluons bind into
LGBs.
In \cite{ouyed01a} we
studied properties of these LGBs. Those
relevant to our
present study are :

i)  The LGBs decay into photons with an
associated lifetime of the order of $10^{-14}$ s.

ii) The mass of the LGBs is of the order of 1 MeV. 

\subsection{Cooling and 2SC layer formation}

The HQS surface layer might enter the 2SC phase  
as illustrated in Figure 1.
 In the QCD phase diagram (Figure 2), 
 ($\rho_{{\bf B_{0}}}$, $T_{{\bf B_{0}}}$) 
is the critical point beyond
which one re-enters the Quark-Gluon-Plasma (QGP) phase
- illustrating the extent of the 2SC layer into the star. 
The star consists of a QGP phase surrounded by
a 2SC layer where the photons (from
the LGB/photon decay) leaking from the
surface of the star 
provides the dominant cooling
source. 
This picture, as illustrated in Figure 2, is only valid 
if neutrino cooling in the 2SC phase is heavily
suppressed as to become slower than the
photon cooling.  
Unfortunately, 
the details of
neutrino cooling in the 2SC phase is still a topic
of debate and studies (\cite{carter00,schaab00}
to cite only few). One can only assume such a scenario which 
 constitutes  the second major assumption
in our model.
In \S \ref{cooling},
we discuss the remaining alternative when
photon cooling is dwarfed by neutrino cooling.

\subsection{LGB decay and photon thermalization}

The photons from LGB decay are generated
at energy $E_{\gamma} < T_{c}$ 
and find themselves immersed in a 
degenerate quark gas. They quickly gain energy via
the inverse Compton process and become thermalized
to $T_{c}$. We estimate the photon mean free path to be smaller
than few hundred Fermi \cite{rybicki79,longair92}
 while the 2SC layer is measured in meters (see
\S 4.2). A local thermodynamic equilibrium is thus reached
with the photon luminosity given by that of  a black body radiation,
\begin{equation}
L_{\gamma} = 3.23\times 10^{52}\ {\rm ergs\ s}^{-1}\
({R_{HQS}\over 5\ {\rm km}})^{2}
({T_{c} \over 10\ {\rm MeV}})^4\ .
\label{three}
\end{equation}
The energy for a single 2SC/LGBs event is thus
\begin{equation}
\Delta E_{LGB}  = \delta_{LGB} M_{2SC}c^2,
\label{four}
\end{equation}
where $M_{2SC} = \delta_{2SC} M_{HQS}$ 
is the  portion of the star in 2SC. Here,
$\delta_{2SC}$ depends on the star's mass
while $\delta_{LGB}$  represent
the portion of the 2SC that is in LGBs (intrinsic
property of 2SC; see \cite{ouyed01a}).  
The photon emission/cooling
 time is then 
\begin{equation}
\Delta t_{cool} = {\epsilon M_{HQS}c^2\over L_{\gamma}}\ ,
\label{five}
\end{equation}
with $\epsilon = \delta_{2SC}\delta_{LGB}$.

\section{Powering Gamma-Ray Bursts}

\subsection{Fireball and baryon loading}

The fireball stems from the LGB
decay and photon thermalization.
The photons  are emitted from the star's surface into 
the vacuum region beneath the inner crust ($\sim 10^{3}$ fm in
size). Photon-photon interaction occurs in a much longer
time than the vacuum region crossing time. Also, the cross-section
for the creation of pairs through interactions with
the electrons in the vacuum region is negligible
(\cite{rybicki79,longair92}).  The
fireball energy is thus directly deposited in the crust.  If its energy
density, $a T_{c}^4$ (with $a$ being the radiation density constant),
exceeds that of the gravitational energy density
in the crust, energy outflow in the form
of ions occurs. 
More specifically, it is the energy transfer from
photons to electrons which drag the positively charged
nuclei in the process.
One can show  that
the condition
\begin{equation}
a T_{c}^4 > {G M_{HQS}\over R_{HQS}} \rho_{crust}\ ,
\label{seven}
\end{equation}
where $\rho_{crust}$ is the crust density
and $G$ the gravitational constant, is equivalent to
\begin{equation}
({T_{c}\over 30\ {\rm Mev}})^4 >
({M_{HQS}\over M_{\odot}})
({5\ {\rm km}\over R_{HQS}})
({\rho_{crust}\over \rho_{drip}})\ ,
\label{eight}
\end{equation}
which is always true if $T_{c} > 30$ MeV.
The  fireball is thus 
loaded with  nuclei
present in the crust.
Note that
the 2SC layer is not carried away during the two-photon decay process
because of the star's high gravitational energy density:
$\rho_{HQS}/ \rho_{drip}>> 1$.

\begin{figure}[t!]
\begin{center}
\epsfig{file=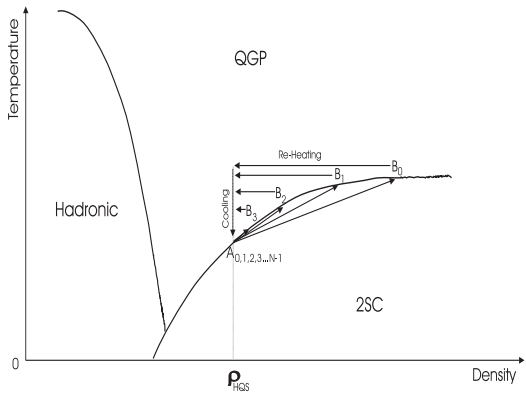,width=3.5in,height=3.5in,angle=-90}
\vskip -1.0in
\caption{
The episodic emission
as illustrated in the QCD phase diagram. 
The 2SC front spreads deep inside the star
and stops at  ${\bf B_{0}}$ before re-entering the
QGP phase. Following photon cooling,  
heat  flows from the core and
re-heats the surface. 
The star then starts cooling until
${\bf A_{1}}$ is reached at which
point  the stage is  set for the 2SC/LGB/photon
 process to start all over again
(${\bf A_{1}}\rightarrow {\bf B_{1}}$)
resulting in another emission.}
\label{figure2}
\end{center}
\end{figure}

\subsection{Episodic behavior}

The star's surface pressure is reduced following photon emission\footnote{The
pressure gradient in the 2SC layer is $\Delta p \propto
(8-5) T_{c}^4$ (\cite{farhi84}) where the
massless gluons (3 out of 8) have
been consumed by the LGB/photon process.}. 
A heat  and mass flux  is thus triggered from the QGP phase
to the 2SC layer  re-heating (above $T_{c}$) and destroying
 the superconductive phase.
The entire star is now  in a QGP phase (5 gluons$\rightarrow$ 8 gluons
at the surface) and
hence the cooling process can start  again.
This corresponds to the transition
$[\rho ({\bf B_{0}}),T ({\bf B_{0}})]\rightarrow 
[\rho_{HQS}, T ({\bf B_{0}})]$
 in the QCD phase diagram (thermal adjustment).
  The stage is  now set
for the 2SC/LGB/photon process to start all over again
resulting in another emission.
For the subsequent emission, however,
we expect the system to evolve to point ${\bf B_{1}}$
generally located at different densities
 and temperatures than ${\bf B_{0}}$.
 The cycle ends after $N$ emissions when 
$\rho({\bf B_{N}})\simeq \rho_{\rm HQS}$.

The time it takes to consume most of the star (the glue
component) by this process is
\begin{equation}
t_{engine}\simeq   {M_{HQS}c^2\over L_{\gamma}} \simeq 
1\ {\rm s}\  ({M_{HQS}\over
M_{\odot}}) ({5\ {\rm km}\over R_{HQS}})^2
({30\ {\rm MeV}\over T_{c}})^4\ ,
\label{totaltime}
\end{equation}
which is representative of the engine's activity.
The above assumes quick adjustment of
the star following each event, but is not necessarily the case for the most
massive stars.

\subsection{Multiple shell emission}

The episodic behavior of the star 
together with the resulting loaded
fireball (we call shell) offers a natural mechanism
for multiple shell emission if 
$T_{c} < 30$ MeV.
Indeed from eq(\ref{eight}) a higher $T_{c}$ value  would imply
extraction of the entire crust in a single emission
and no loading of the subsequent fireballs.
Clearly, $T_{c} < 30$ MeV must be
considered if multiple ejections are to
occur\footnote{Even if $T_c$ 
turns out to be greater than 30 MeV, in which case the
entire crust will be blown away (eq(\ref{eight})),
one can imagine mechanisms where crust material is replenished.
By accretion, for instance, if the HQS is part of a binary.
There are also geometrical considerations where
asymmetric emission/ejection can occur
due to the rapid rotation of Quark Stars; here
only a portion of the crust is extracted at a
time.
This aspect of the model requires better knowledge of the
conditions and environments where HQSs are formed.}.

The fraction ($f$) of the crust 
 extracted in a single event is, 
\begin{equation}
\Delta M_{crust} = f M_{crust}\ .
\label{nine}
\end{equation}
The  shell is accelerated
with the rest of the fireball converting most
of the radiation energy into bulk kinetic energy.
The corresponding Lorentz factor we estimate to be,
\begin{equation}
\Gamma_{\rm shell} \simeq
{\epsilon M_{HQS}\over f M_{crust}}\ ,
\label{ten}
\end{equation}
where we used eq(\ref{four}) and eq(\ref{nine}).
$\epsilon$ and $f$ depend on the star's mass
and characterize the two emission regimes
in our model.

\section{The two regimes}

When the inner crust density is the neutron drip value, one finds
a minimum mass star of $\sim 0.015M_{\odot}$. For 
masses above this critical value, the corresponding
 crusts are thin and light.
They do not exceed few kilometers in thickness.
Matter at the density of such crusts
is a Coulomb lattice of iron and nickel
all the way from the inner edge to the surface of the
star (\cite{baym71}). For masses below
$0.015M_{\odot}$, the crust can extend up to thousands of kilometers with
densities much below the neutron drip.
This allows us to identify two distinct emission regimes
for a given $T_{c}$ ($< 30$ MeV).

\subsection{Light stars ($M_{HQS} < 0.015M_{\odot}$)}

These are objects whose average density is $\sim \rho_{HQS}$
($M_{HQS} \simeq {4\pi\over 3}R_{HQS}^{3}\rho_{HQS}$).
The 2SC front extends deeper inside
the star ($\delta_{2SC}\sim 1$).
The star can be represented by a system close
to ${\bf A_{0}}$ in Figures 2.
Each of the few emissions (defined by $\epsilon$)
is thus capable of consuming a big portion of the star.
Furthermore, the entire crust  material
can be extracted in a few 2SC/LGB/photon cycles
($\rho_{crust}/\rho_{drip} \ll 1$).

Using eq(\ref{totaltime}), the few emissions lead to  
\begin{eqnarray}
t_{tot} &\simeq& {\rm fraction}\times t_{engine}\nonumber\\
&\simeq& {\rm fraction}\times 0.25\ {\rm s}\ 
({M_{HQS}\over 0.01\ M_{\odot}})
({1\ {\rm km}\over R_{HQS}})^2
({30\ {\rm MeV}\over T_{c}})^4\ ,\nonumber\\
\label{tsmall}
\end{eqnarray}
where $t_{tot}$ is representative of the
observable time which takes into account
the presence of the crust. 

\subsection{Massive stars ($M_{HQS} \ge 0.015M_{\odot}$)}

The surface density of a massive star  being
 that of a light star, defines 
a standard unit in our model. 
In other words, the  mass of the 2SC layer
in a massive star case is
\begin{equation}
\Delta M_{2SC,m}\simeq M_{2SC,l}\ ,
\label{twelve}
\end{equation}
where ``$m$'' and ``$l$'' stand for massive and light,
respectively. It implies
\begin{equation}
{\Delta R_{2SC,m}\over R_{2SC,m}} \simeq
{1\over 3} ({R_{l}\over R_{m}})^3
\simeq {1\over 3} ({1\ {\rm km}\over 5\ {\rm km}})^3
\simeq 0.003\ . 
\label{thirteen}
\end{equation}
For a typical star of 5 km in radius,
we then estimate a 2SC layer of about
15 meters thick (much larger than the photon mean free path thus justifying
the local thermal equilibrium hypothesis). Equivalently,
\begin{equation}
\bar{\epsilon} = {M_{l}\over M_{m}} \simeq 
  ({1\ {\rm km}\over 5\ {\rm
km}})^3 \simeq 0.01\ ,
\label{fourteen}
\end{equation}
where $\bar{\epsilon}$ is the average value.
This naturally account for many events
(or $N$ fireballs). The average number
of fireballs with which an entire star is consumed is thus 
\begin{equation}
N\simeq {1\over \bar{\epsilon}}\simeq 100\ .
\label{N}
\end{equation}
Since most of the crust is at densities close to the
neutron drip value, eq(\ref{eight}) implies that only a tiny
part of the crust surface material (where $\rho_{crust}
<< \rho_{drip}$) can be extracted by each of the fireballs.
This allows for a continuous loading of the fireballs.

The total observable time in our simplified approach
is thus, 
\begin{equation}
t_{tot}\simeq t_{engine}
= 1\ {\rm s}\, ({M_{HQS}\over M_{\odot}})
({5\ {\rm km}\over R_{HQS}})^2
({30\ {\rm MeV}\over T_{c}})^4\ .
\label{tmassive}
\end{equation}
\newline

We  isolated two regimes:
\newline

(i) Light stars $\Rightarrow$ short  emissions.
\newline

(ii) Massive stars $\Rightarrow$ long emissions.
\newline

It appears, according to BATSE
(Burst and Transient Source Experiment
 detector on the COMPTON-GRO satellite), that the bursts can be
classified into two distinct categories: short ($< 2$ s) bursts
 and long ($> 2$ s, typically $\sim 50$ s) bursts.
The black body behavior
($T_{c}^4$) inherent to our
model puts stringent constraints on the value
of $T_{c}$ which best comply with these observations. 
Using $T_{c}\simeq 10$ MeV, from eq(\ref{tsmall})
and eq(\ref{tmassive})
 we obtain in the star's rest frame
\begin{equation}
t_{tot} \simeq  81\ {\rm s}\ 
 ({M_{HQS}\over M_{\odot}})
({5\ {\rm km}\over R_{HQS}})^2\ ,
\label{seventeen}
\end{equation}
for massive stars (suggestive of long  GRBs), and
\begin{equation}
t_{tot} \simeq   2\ {\rm s}\ 
({M_{HQS}\over 0.01\ M_{\odot}})
({1\ {\rm km}\over R_{HQS}})^2\ ,
\label{eighteen}
\end{equation}
for light stars (suggestive of short GRBs).
There is
a clear correlation (almost one to one)
between the observed burst time and the
time at which the source ejected the specific shell
(see Figure 3 in \cite{kobayashi97},
for example). Note that $T_{c} \simeq 10\ {\rm MeV}$ implies that
only a portion of the crust is extracted. This 
  is also consistent with our previous assumption
($T_{c} < 30$ MeV) and subsequent calculations.

Eq(\ref{seventeen}) and eq(\ref{eighteen}) is simply eq(\ref{totaltime})
rescaled  to the appropriate object size. 
We separated two regimes due to intrinsic
differences in the engine and the crust.
From the engine point of view, 
massive stars generate many more emissions when compared
to light ones, and no substantial
reduction of the engine time is expected because
of the omni-presence of the crust.
Another important difference  is 
related to the physics of the multiple re-adjustments
 following each event  which is more 
pronounced for very massive stars.
The latter among other
factors is related to $\epsilon$ which can vary from
one event to another.

\section{Features and predictions}

\subsection{GRB energies}

The maximum available
energy is when the heaviest HQS
($M_{HQS,max}\simeq 2 M_{\odot}$) is entirely consumed.
That is,
\begin{equation}
E_{LGB,max} \simeq 4 \times
10^{54}\ {\rm ergs}\ .
\label{twentythree}
\end{equation}
The corresponding GRB energy is 
\begin{equation}
E_{GRB,max} \simeq 1.6 \times
10^{54}\ {\rm ergs}\ ,
\label{twentyfour}
\end{equation}
where we used a fiducial conversion
efficiency of 40\%.

Since 
$M_{HQS,min} < 0.015M_{\odot}$  we conclude that,
\begin{equation}
E_{LGB,min} < 3 \times 10^{52}\ {\rm ergs}\ ,
\label{twentythree2}
\end{equation}
implying
\begin{equation}
E_{GRB,min} < 1.2 \times
10^{52}\ {\rm ergs}\ .
\label{twentyfour2}
\end{equation}

\subsection{GRB total duration}

From eq(\ref{seventeen}) and eq(\ref{eighteen}) we have
\begin{equation}
t_{tot} \simeq  81\ {\rm s}\ ,
\label{twentyfive}
\end{equation}
for typical massive stars, and
\begin{equation}
t_{tot} \simeq   2\ {\rm s}\ ,
\label{twentysix}
\end{equation}
for typical light stars. 

Our estimate of the duration time for the massive star case
should be taken as a lower limit. As we have said,
a complete model should take into account star
re-adjustments. Nevertheless, we can still
account for a wide range in GRB duration  by
an appropriate choice of different values of the mass and radius.

\section{Discussion and Conclusion}
\label{discuss}

\subsection{Existence and formation of HQSs}

In the last few years, thanks to the large amount of fresh
observational data collected by the new generation of X-ray and
$\gamma$-ray satellites, new observations suggest that the compact
objects associated with the X-ray pulsars, the X-ray bursters,
particularly the SAX J1808.4-3658, are good quark stars candidates (see
Bombaci in this volume and \cite{li99}).
If one assumes that these plausible quark stars form via
the ``standard'' supernova mechanism or by conversion of neutron stars
then the  two regimes (Heavy 
and Light stars) 
discussed in our model are difficult to
account for. 
It has been argued however that
quark stars formation mechanisms may be numerous 
and ``exotic'' 
(early  discussions can be found in
\cite{alpar87,glendenning97}).
In the case of 4U 1728-34
(where a mass of less than $1.0M_{\odot}$ was derived; \cite{bombaci00}),
it  seems that
accretion-induced collapse of white dwarfs is a favored 
formation mechanism. If the
quark star formed via the direct conversion mechanism then it
required too much mass (at least $\sim 0.8M_{\odot}$ to be ejected
during the conversion). 
Other formation scenarios are discussed 
in \cite{hong01,ouyed01c}.

\subsection{Neutrino cooling and HQSs}
\label{cooling}

If it turns out that neutrino cooling
is still very efficient in the 2SC phase,
one must consider the
scenario where the entire HQS enters the 2SC phase
(for comparison of cooling
paths between quark stars and neutron stars
and the plausible effects of 2SC on cooling
we refer the interested reader to \cite{schaab00,blaschke00,blaschke01}).
Here, the 2SC/LGB/photon process
(the fireball) occurs only once and inside the entire star. 
Furthermore, one must 
involve more complicated physics (such as that of the
crust) to account for the
episodic emissions so crucial to any model
of GRBs. It is not clear at the moment how to achieve this goal
and is left as an avenue
for future research.

\subsection{2SC-II stars}

 The 2SC/LGB/photon process might proceed until
one is left with an object made entirely of 2SC. We name such
objects  {\it 2SC-II} stars\footnote{The ``II'' in 2SC 
is a simple reminder of the final state of the
star, namely the 2SC with only 5 gluons.} which are still
bound by strong interactions (their density is constant
$\sim \rho_{HQS}$).
2SC-II stars carry 
an Iron/Nickel crust left over from the GRB phase. The crust
mass range is $0 <M_{2SC,crust}<5\times 10^{-5}M_{\odot}$
 depending on the efficiency of crust extraction/ejection during the GRB
phase.

BATSE  observes on average one burst per day. This corresponds,
with the simplest model - assuming no cosmic evolution of the rate - to 
about once per million years in a galaxy \cite{piran99a}. 
In the Milky way we thus expect up to $10^{5}$ of 2SC-II stars.
Nevertheless, they are tiny enough ($M\le
10^{-2}M_{\odot}$, $R\le 1$ km) to be difficult to detect.

\subsection{The Mass$-$Radius Plane}

Take observed GRBs with
known energies and total duration.  
From the  burst total energy  $E_{GRB}
\simeq 0.4 Mc^2$ we derive the mass while
 the burst total duration ($t_{tot}$) 
gives us the radius (using eq(\ref{tmassive})
with $T_{c}\simeq 10$ MeV).
In Figure 3 we plot the resulting Mass$-$Radius.
The solid curve shows the
$M_{HQS}={4\pi\over 3}\rho_{HQS}R_{HQS}^3$ equation which is a reasonable
approximation for HQSs. 
While the GRB data set used is limited nevertheless
it seems to support the idea that extremely compact objects ($M\propto
R^3$) are behind GRBs activity within our model.

\begin{figure}[t!]
\begin{center}
\epsfig{file=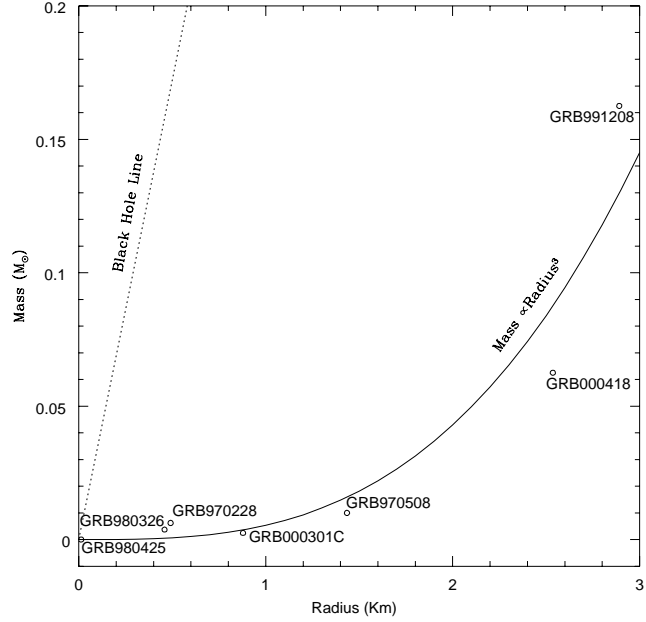,height=3.5in}
\caption{The Mass$-$Radius
plane derived in our model using few existing GRBs with known
energies and total duration.  The solid curve shows the
$M_{HQS}={4\pi\over 3}\rho_{HQS}R_{HQS}^3$ equation for $\rho_{HQS} \simeq
9\rho_{N}$.}
\label{figure3}
\end{center}
\end{figure}

\end{document}